\documentclass[pra,aps,twocolumn,nobibnotes,nofootinbib]{revtex4}
\usepackage{amsmath}
\usepackage{amssymb}
\usepackage{bm}
\usepackage{epsfig}
\usepackage{graphicx}
\begin{document}

\title{Effect of structure anisotropy on low temperature spin dynamics in quantum wells}

\author{M.~M.~Glazov}
\email{glazov@coherent.ioffe.ru} \affiliation{A. F. Ioffe
Physico-Technical Institute, Russian Academy of Sciences, 194021
St.~Petersburg, Russia}

\begin{abstract}
Spin dynamics of two-dimensional electron gas confined in an
asymmetrical quantum well is studied theoretically in the regime
where the scattering frequency is comparable with the spin
precession frequency due to the conduction band spin splitting.
The spin polarization is shown to demonstrate quantum beats. If
the spin splitting is determined by both bulk and structural
asymmetry mechanisms the beats are damped at zero temperature even
in the absence of a scattering. We calculate the decay of spin
beats due to the thermal broadening of the electron distribution
function and electron scattering. The magnetic field applied along
the structure growth axis is shown to increase the frequency of
the beats and shift system towards the collision dominated regime.
\end{abstract}

\date{7 February 2007} \maketitle

\section{Introduction}

The spin dynamics in semiconductor nanostructures is a field of
topical interest~\cite{zutic:323}. The progress in the
nanostructure technology allows to fabricate semiconductor quantum
wells (QWs) of high finesse where the carrier scattering time
$\tau$ caused by the disorder amounts to several tens of
picoseconds. The conduction band spin splitting $\Delta$ in such
structures can reach several meV which implies $\Delta\tau/\hbar
\gtrsim 1$. In this case the spin beats are observed in
time-resolved experiments for low enough temperatures where
electron scattering role is diminished~\cite{brand02}.

Theoretically, these spin beats were studied in
Ref.~\cite{gridnev01}, where the consideration was restricted to
the case of isotropic conduction spin splitting, see also
Ref.~\cite{wu04diff}; some results for anisotropic splitting were
given in Ref.~\cite{culcer06}. The regime of isotropic spin
splitting being comparable with the electron Fermi energy was
discussed in Ref.~\cite{grimaldi05}. However, the evidence is
growing that the spin splitting in the state-of-the-art QW samples
is anisotropic function of the electron
wavevector~\cite{averkiev06,giglberger06}. Therefore the full
interpretation of the experimental data on low temperature spin
dynamics in high-quality QW samples needs a theory which takes
into account the anisotropy of the spin splitting. The goal of the
present paper is to put forward the theory of spin beats for the
case of arbitrary relation between the spin splitting and the
collisional broadening and for the arbitrary anisotropy of the
spin splitting. In Sec.~\ref{sec:kin} we present the kinetic
equation for spin distribution function and discuss the dephasing
of the beats due to thermal broadening of the electron
distribution and due to the anisotropy of the spin splitting.
Section~\ref{sec:scatter} is devoted to the decay of spin beats
due to the scattering, and the effect of an external magnetic
field on the spin beats is discussed in Sec.~\ref{sec:magnetic}.

\section{\label{sec:kin} Kinetic equation and spin dynamics in the absence of the
scattering}

We consider a QW grown along $z\parallel [001]$ from zinc-blende
lattice semiconductor. The spin dynamics of the two-dimensional
electron gas in the absence of magnetic field is determined from
the kinetic
equation~\cite{dyakonov72,dyakonov86,ivchenko88,glazov04a}
\begin{equation}\label{kin}
\frac{\partial \bm s_{\bm k}}{\partial t} + \bm s_{\bm k} \times
\bm \Omega_{\bm k} + \bm Q_{\bm k}\{\bm s_{\bm k}\} =0,
\end{equation}
where $\bm s_{\bm k}$ is the average spin of the electron in the
state with the wavevector $\bm k$, i.e. $\bm s_{\bm k}$ is the
spin distribution function, $\bm \Omega_{\bm k}$ is the spin
precession frequency originated from the $\bm k$-linear terms in
the electron effective
Hamiltonian~\cite{bychkov84,dyakonov86,ivchenko05a}
\begin{equation}\label{omega_k}
\mathcal H_{SO} = \frac{\hbar}{2}\ ({\bm \sigma}\cdot {\bm
\Omega}_{\bm k}) = \beta_1 (\sigma_y k_y - \sigma_x k_x) + \beta_2
(\sigma_x k_y - \sigma_y k_x)\:,
\end{equation}
with $\beta_1$ and $\beta_2$ being Bulk Inversion Asymmetry (BIA
or Dresselhaus) and Stucture Inversion Asymmetry (SIA or Rashba)
spin splitting constants~\footnote{We ignore the $k^3$ terms in
the effective Hamiltonian assuming that the carrier density is not
too high. The Interface Inversion Asymmetry term (IIA) has the
same symmetry as BIA term thus $\beta_1$ can be considered as a
sum of BIA and IIA spin splitting constants.}, and $\bm Q_{\bm
k}\{\bm s_{\bm k}\}$ is the collision integral. Here we use the
coordinate frame with $x\parallel [100]$, $y\parallel [010]$. From
now on we consider spin-independent isotropic scattering. We
assume that the average spin splitting on the Fermi surface
$\Delta = \hbar \bar{\Omega}_{k_F}$ (where $k_F$ is the Fermi
wavevector, $\bar{\Omega}_{k_F}$ is the angular average of $|\bm
\Omega_{\bm k}|$) is much smaller than electron Fermi energy
$E_F$. It allows us to avoid the antisymmetrization of the
collision integral~\cite{ivchenko89,grimaldi05} and use the simple
expression
\begin{equation}\label{Qk}
\bm Q_{\bm k}\{\bm s_{\bm k}\} =\frac{1}{\tau}\left[\bm s_{\bm k}
- \bar{\bm s}_k\right],
\end{equation}
where $ \bar{\bm s}_k = (2\pi)^{-1}\int_0^{2\pi} \bm s_{\bm
k}\mathrm d \varphi_{\bm k} $ is the angular-averaged spin
distribution function, $\varphi_{\bm k}$ is the axial angle of the
wavevector $\bm k$ and $\tau^{-1}$ is the scattering rate. We note
that the kinetic equation \eqref{kin} itself is applicable for
$E_F\tau \gg \hbar$.

Our goal is to solve Eq.~\eqref{kin} for the arbitrary relation
between spin precession frequency $\bm \Omega_{\bm k}$ and the
scattering rate $\tau^{-1}$. We assume that at $t=0$ the
quasi-equilibrium distribution of spin $z$-component is
created~\cite{dyakonov72,glazov04a}:
\[
s_{z,\bm k}(t=0) = \frac{1}{2}(f_{\bm k, 1/2} - f_{\bm k, -1/2}),
\]
where $f_{\bm k, s} = \{\exp{[(\mu_s - E_{\bm k})/k_B T]} +1
\}^{-1}$ is the Fermi-Dirac distribution function, $\mu_{\pm 1/2}$
are the chemical potentials for electrons with spin $z$-components
$\pm 1/2$ respectively, $E_{\bm k}$ is electron's dispersion and
$k_B T$ is the temperature measured in energy units. Under the
assumption of low spin polarization,
\[
P_s = \frac{\sum_{\bm k} (f_{\bm k, 1/2} - f_{\bm k,
-1/2})}{\sum_{\bm k} (f_{\bm k, 1/2} + f_{\bm k, -1/2})} \ll 1,
\]
the initial spin distribution reduces to
\begin{equation}\label{szinit}
s_{z, \bm k} (0) = \frac{P_s E_F}{k_B T} f_{\bm k} (1-f_{\bm k}),
\end{equation}
where $f_{\bm k}$ is the Fermi-Dirac distribution function with
$\mu = (\mu_{1/2} + \mu_{-1/2})/2$. For simplicity the temperature
is taken to be zero, $T=0$, the effects of non-zero temperature
will be briefly discussed below.

First, we consider a limit of a clean system, where $\tau\to
\infty$. Thus, we neglect the collision term in Eq.~\eqref{kin}.
Then, the spin of the electron with the wavevector $\bm k$
precesses around $\bm \Omega_{\bm k}$ according to
\begin{equation}\label{skt}
s_{z,\bm k} (t) = s_{z,\bm k}(0) \cos{\Omega_{\bm k}t}.
\end{equation}
The total spin polarization can be represented as average of Eq.
\eqref{skt} over the initial spin distribution, which in the limit
of $T=0$ can be replaced by an average over the axial angle
$\varphi_{\bm k}$ of the electron wavevector:
\begin{equation}\label{stott}
s_{z} (t) = s_z(0) \int \frac{\mathrm d \varphi_{\bm k}}{2\pi}
\cos{\left({\frac{2 k}{\hbar} \sqrt{\beta_1^2+\beta_2^2 -
2\beta_1\beta_2 \sin{2\varphi_{\bm k}}}t}\right)}.
\end{equation}

In the limiting case of the \emph{isotropic spin splitting} where
$\beta_1=0$ or $\beta_2=0$, i.e. either BIA or SIA term is present
in the spectrum only, the absolute value of the vector $\bm
\Omega_{\bm k}$ is $\varphi_{\bm k}$ independent and according to
Eq. \eqref{stott}
\[
s_{z} (t) = s_{z}(0) \cos{\Omega_{k_F}t},
\]
where $k_F$ is the Fermi wave vector. One can see that the total
spin oscillates and returns exactly to the initial value each
period $2\pi/\Omega_{k_F}$~\cite{gridnev01}.

The spin splitting anisotropy leads to a difference of the spin
precession frequencies in different points of the $\bm k$ space.
As the precession frequencies are not commensurable the spin will
never return to the initial value, however, its time decay will be
very slow. For example, in the limiting case $\beta_1 =
\beta_2\equiv\beta$ (which corresponds to the \emph{strongest spin
splitting anisotropy}) the angular integration in Eq.
\eqref{stott} can be easily carried out with the
result~\cite{culcer06}
\begin{equation}\label{stot_aniso}
s_{z} (t) = s_{z}(0) \mathrm J_0\left(\frac{4\beta k_F
t}{\hbar}\right),
\end{equation}
where $\mathrm J_0(x)$ is the Bessel function. At $\beta k_F
t/\hbar \gg 1$ the spin oscillates and decays as $t^{-1/2}$. Such
an asymptotics ($t^{-1/2}$) holds true even in the case of the
arbitrary ratio of $\beta_1$ and $\beta_2$ provided
$\sqrt{|\beta_1\beta_2|}t/\hbar \gg 1$, the latter can be seen
directly from Eq. \eqref{stott} since the main contribution to the
integral comes from the points of the stationary phase,
$\varphi_{\bm k} = \pi/2$, $3\pi/2$. We note that this relaxation
is `reversible' as it is caused by the spread of the precession
frequencies, like for the spin dephasing induced by the spread of
$g$-factor values.

In the end of this Section we shortly comment on the effect of
non-zero temperature on the smearing out of the spin beats. For
the simplest case of \emph{isotropic spin splitting} the beats are
damped at $k_B T \ll E_F$ as
\begin{equation}\label{t-damp}
s_z(t) = s_z(0) \frac{\pi \alpha t}{\sinh{(\pi \alpha t)}}
\cos{\Omega_{k_F} t}\,,
\end{equation}
where $\alpha = \Omega_{k_F} k_B T/2E_F$. On a long time-scale
($\alpha t\gg 1$) the polarization oscillations decay
exponentially with the time constant
\begin{equation}\label{tau-t}
\tau_T = \frac{2E_F}{\pi \Omega_{k_F} k_B T} \,.
\end{equation}
We note that the exponential time decay of the spin beats takes
place in the regime of \emph{anisotropic spin splitting} as well,
the time constant has the same order of the magnitude as $\tau_T$.
If the initial spin distribution has an energy width $\Delta E\gg
k_B T$ the beats will be damped during the time $\sim
E_F/(\Omega_{k_F}\Delta E)$, provided this time is shorter than
the energy relaxation time.

\section{Effect of scattering}\label{sec:scatter}

Now we are in the position to discuss effect of the electron
scattering on the spin beats. First we solve Eq. \eqref{kin} for
the case of $T=0$ and afterwards we discuss the competition
between the damping of the beats due to the scattering and due to
the thermal broadening of the distribution function. In order to
solve Eq. \eqref{kin} we rewrite it in components as
\begin{eqnarray}
\frac{\partial s_{x,\bm k}}{\partial t} - \Omega_{y,\bm
k}s_{z,\bm k} + \frac{s_{x,\bm k}}{\tau} =0, \label{kin1} \\
\frac{\partial s_{y,\bm k}}{\partial t} + \Omega_{x,\bm
k}s_{z,\bm k} + \frac{s_{y,\bm k}}{\tau} =0, \label{kkin1} \\
\frac{\partial s_{z,\bm k}}{\partial t} + \Omega_{y,\bm k}
s_{x,\bm k}  - \Omega_{x,\bm k} s_{y,\bm k} + \frac{s_{z,\bm k} -
\bar{s}_{z,k}}{\tau} =0.\label{kin1last}
\end{eqnarray}
In deriving Eqs. \eqref{kin1} we made use of the fact that $z$ is
the principal axis of the system, i.e. the relaxation of the spin
parallel to $z$ will not lead to the average in-plane spin
polarization of the carriers. It follows from Eqs. \eqref{kin1},
\eqref{kkin1} that
\[
\left(\frac{\partial }{\partial t} +\frac{1}{\tau}\right)
\bigl(\Omega_{y,\bm k} s_{x,\bm k}  - \Omega_{x,\bm k}s_{y,\bm k}
\bigr) - \Omega_{\bm k}^2 s_{z,\bm k}  =0,
\]
or
\begin{equation}\label{szfin}
\left(\frac{\partial }{\partial t} +\frac{1}{\tau}\right)
\frac{\partial }{\partial t} s_{z,\bm k} + \Omega_{\bm k}^2
s_{z,\bm k} + \left(\frac{\partial }{\partial t}
+\frac{1}{\tau}\right) \frac{s_{z,\bm k} - \bar{s}_{z}}{\tau} = 0.
\end{equation}
The initial conditions for Eq. \eqref{szfin} are $s_z(t=0)=s_z(0)$
and $\partial s_z(t=0)/\partial t=0$ which follows from Eq.
\eqref{kin1last}.

In the \emph{isotropic case} the spin precession frequency
$\Omega_{\bm k}$ is angular independent and the last term of Eq.
\eqref{szfin} vanishes. The spin dynamics is then described by:
\begin{equation}\label{gridnevNEW}
s_z(t)=s_z(0)\left[\cosh{\left(\frac{qt}{2\tau}\right)} +
\frac{1}{q}\sinh{\left(\frac{qt}{2\tau}\right)}\right]\mathrm
e^{-\frac{t}{2\tau}},
\end{equation}
where $q=\sqrt{1-4\Omega_{k_F}^2\tau^2}$, in agreement with
Ref.~\cite{gridnev01}. Note that, in Ref.~\cite{grimaldi05} this
expression [Eq. (17)] is presented with an error: the second term
in brackets is missing and the condition $\partial
s_z(t=0)/\partial t=0$ is violated.

According to Eq. \eqref{gridnevNEW} one can identify two
qualitatively different regimes of the spin relaxation: (i) spin
precession regime with the exponentially decaying spin beats at a
frequency $\sqrt{\Omega_{k_F}^2-(2/\tau)^{2}}$ and the decay time
constant $\tau_B = 2\tau$ and (ii) the collision dominated spin
relaxation regime showing exponential damping of the total spin
with the spin relaxation time $\tau_{DP} =
1/(\Omega_{k_F}^2\tau)$. The transition between the two regimes
takes place at $\Omega_{k_F}\tau=1/2$.

In the general case, where both BIA and SIA terms are present, the
spin precession frequency $\Omega_{\bm k}$ is a function of
$\varphi_{\bm k}$ and all angular harmonics of $s_{z,\bm k}$ are
intermixed. Eq. \eqref{szfin} can be reduced to a set of linear
algebraic equations by the decompositon of $s_{z,\bm k}$ into
angular harmonics as follows
\[
s_{z,\bm k} = \sum_{m=-\infty}^{\infty} s_{z,m} \mathrm e^{\mathrm
i m \varphi_{\bm k}}.
\]
Eq. \eqref{szfin} is thus equivalent to an infinite system of the
linear equations for the values $s_m(t)$ which reads
\begin{widetext}
\begin{equation}\label{szfin_m}
\left(\frac{\partial }{\partial t} +\frac{1}{\tau}\right)
\frac{\partial s_{z,m}}{\partial t}  + \left(\frac{\partial
}{\partial t} +\frac{1}{\tau}\right)
\frac{1-\delta_{m,0}}{\tau}s_{z,m}  + \tilde \Omega_0^2 s_m +
\mathrm i \tilde \Omega_2^2 s_{z,m-2} - \mathrm i  \tilde
\Omega_{2}^2 s_{z,m+2} = 0.
\end{equation}
\end{widetext}
Here the integer $m$ runs from $-\infty$ to $\infty$,
$\delta_{m,0}$ is the Kronecker symbol and
\begin{equation}\label{tildeomega1}
\tilde \Omega_0^2 = \frac{4k^2}{\hbar^2}
(\beta_1^2+\beta_2^2),\quad  \tilde \Omega_2^2 =
\frac{4k^2}{\hbar^2} \beta_1\beta_2.
\end{equation}
These equations should be supplemented with the initial condition
$s_{z,m}(t=0)= s_z(0)\delta_{m,0}$ and $\partial
s_{z,m}(t=0)/\partial t =0$.

\begin{figure*}[htbp]
  \includegraphics[width=0.9\linewidth]{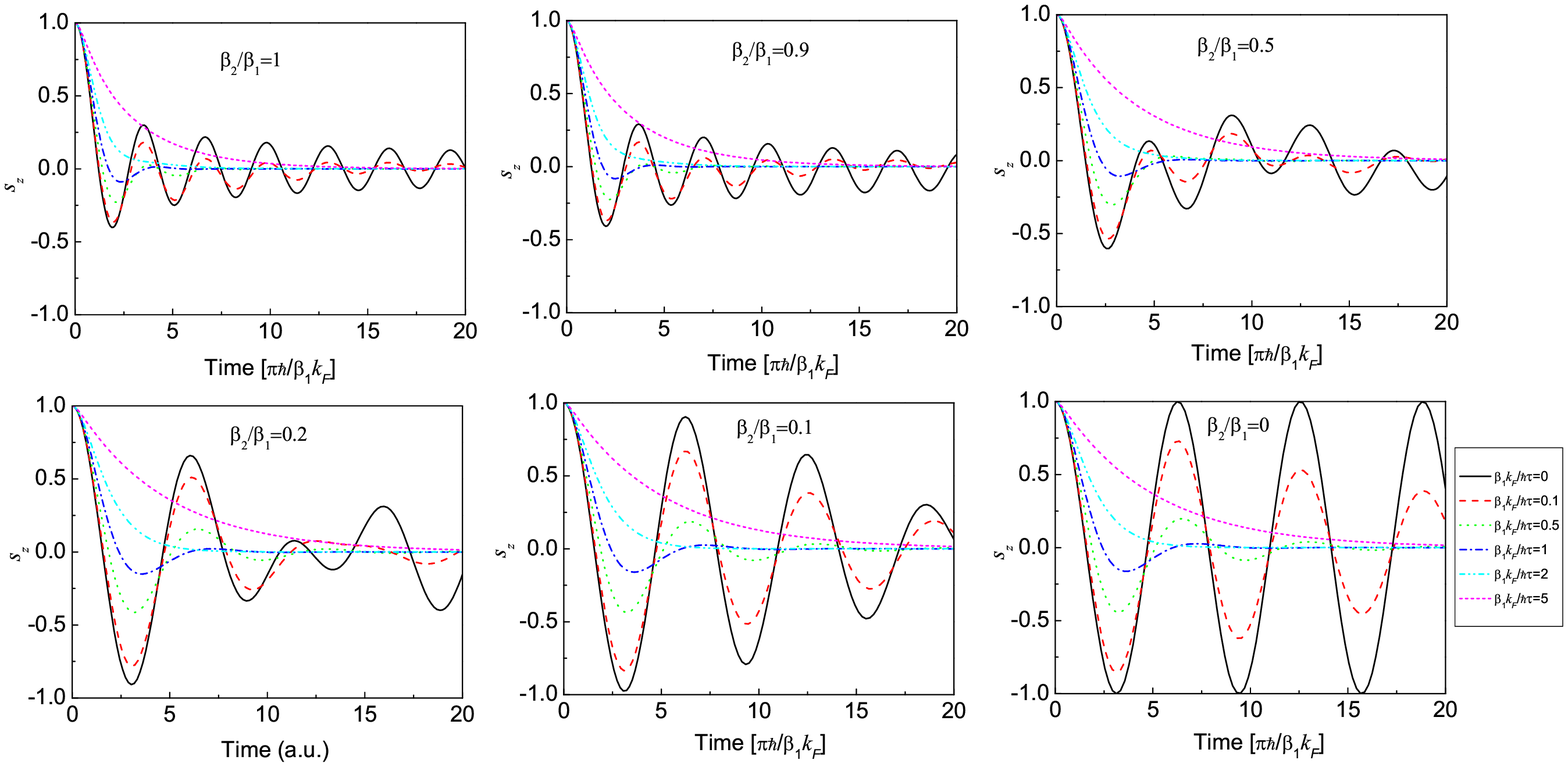}\\
  \caption{The time dynamics of the average $z$-component of the electron
  spin for different ratio of BIA and SIA terms for a fixed $\beta_1$ term.
  The time units are $\pi\hbar/\beta_1 k_F$. The ratio
  $\beta_2/\beta_1$ is shown above each panel. Various
  curves correspond to different scattering rates, see legend.}\label{fig:osc}
\end{figure*}

Figure~\ref{fig:osc} presents the time dependence $s_z(t)$
obtained by the numerical solution of Eq. \eqref{szfin_m}.
Different panels of Fig.~\ref{fig:osc} show the results for
different ratios $\beta_2/\beta_1$ calculated at a fixed
$\beta_1$. Eq. \eqref{tildeomega1} clearly shows that the
dependence $s_z(t)$ is invariant under the replacement $\beta_1
\leftrightarrow \beta_2$. Different curves in each panel are
calculated for different scattering rate $\tau^{-1}$. Even a small
admixture of the other term to the strong first term in
Hamiltonian \eqref{omega_k}, e.g. $\beta_2/\beta_1=0.2$, leads to
the damping of the oscillations and quite complicated behavior of
$s_z(t)$.

The inclusion of the scattering first smears the oscillations. The
damping time is non-monotonous function of $\tau^{-1}$ in
accordance with the qualitative discussion presented above. For
small scattering rates the scattering enhances spin relaxation,
while for the stronger scattering the spin dynamics becomes
collision dominated and the spin relaxation time becomes longer.

The possibility for the experimental observation of these
oscillations depends strongly on the relation between the average
spin precession frequency $\bar \Omega_{k_F}$, scattering time
$\tau$ and thermal broadening time $\tau_T$. Namely, the condition
$\bar \Omega_{k_F} \tau \gtrsim 1/2$ guarantees the spin
precession regime. Moreover, $\bar \Omega_{k_F}\tau_T$ should be
larger than unity in order avoid thermal damping of the
oscillations.

The dominant broadening mechanism is therefore determined by the
shortest of $\tau_T$ and $\tau$. For very low temperatures
$\tau_T$ is very long while $\tau$ is determined by the carrier
scattering on the interface roughness and remote impurities. In
this regime the beats are damped due to electron elastic
scattering. In the regime of the intermediate temperatures where
electron-electron scattering dominates over the elastic processes
governing the momentum relaxation~\cite{leyland06}, the scattering
rate is proportional to $\tau^{-1} \propto k_B^2 T^2/\hbar E_F$
and scales as $T^2$~\cite{glazov04a}, whereas the thermal
broadening rate $\tau_T^{-1} \propto\bar \Omega_{k_F} k_B T/E_F
\propto T$, see Eq. \eqref{tau-t}. Their ratio $\tau_T/\tau$
scales as $k_B T/\hbar \bar \Omega_{k_F}$ and can be both larger
or smaller than $1$. For example, in the experimental conditions
of Ref.~\cite{brand02} $\hbar \bar \Omega_{k_F} \approx 1$~meV and
for the temperatures larger than $\hbar \bar \Omega_{k_F}/k_B
\approx 12$~K the damping of the beats is due to the
electron-electron scattering. We note that in the experiments
quasi-equilibrium initial spin distribution can be established
during the energy relaxation time $\tau_E$ which can be comparable
with $\tau$. In such a case one can not separate two contributions
to the beats decay.

In the steady-state experiments one measures the spin polarization
degree at the continuous spin generation rate. Let us assume that
this rate $\bm G_s$ is directed along $z$-axis and is isotropic in
the $\bm k$ space. Provided the carrier lifetime is much longer
than the spin relaxation time, the stationary solution of Eqs.
\eqref{kin1}-\eqref{kin1last} reads
\begin{equation}\label{sz_steady1}
s_{z,\bm k} = \frac{\bar s_{z,k} + G_s \tau}{1+\Omega_{\bm
k}^2\tau^2},
\end{equation}
where the average spin-$z$ component is given by
\begin{equation}\label{szbar}
\bar s_{z,k} = \frac{G_s\tau}{[(1+\tilde \Omega_0^2\tau^2 + 2
\tilde \Omega_2^2\tau^2)(1+\tilde \Omega_0^2\tau^2 - 2\tilde
\Omega_{2}^2\tau^2)]^{\frac{1}{2}}-1}.
\end{equation}
We remind that $\tilde \Omega_0^2$ and $\tilde \Omega_{2}^2$ are
defined by Eq. \eqref{tildeomega1}. For example, if the spin
splitting is \emph{isotropic} the proportionality coefficient
between average spin $\bar s_z$ and $G_s$ is simply given by
$(\tilde \Omega_0^2\tau)^{-1}$ which coincides with
Dyakonov-Perel' spin relaxation rate in the collision dominated
regime~\cite{dyakonov72,dyakonov86}.

\section{Spin dynamics in a magnetic field}\label{sec:magnetic}

It is well known that the magnetic field $\bm B \parallel z$ slows
down the spin relaxation in the collision dominated
regime~\cite{ivchenko73,glazov04}. The goal of this Section is to
analyze the effects of the magnetic field for the arbitrary
relation between $\bar \Omega_{k_F}$ and $\tau^{-1}$. Here we
neglect the non-zero temperature effects.

The magnetic field has a two-fold effect on the spin dynamics.
First, it causes the precession of electron spins due to Larmor
effect and, second, the cyclotron motion of the electrons in the
field leads to the rotation of the wavevector $\bm k$ of a given
electron and correspondingly the spin precession vector $\bm
\Omega_{\bm k}$. The effects can be taken into account as
additional terms in the left hand side of the kinetic equation
\eqref{kin}, namely, $\bm s_{\bm k}\times \bm \omega_{L}$ and
$\hat \Lambda \bm s_{\bm k} =\bm \omega_{C}[\bm k \times
\partial \bm s/\partial \bm k]$ with $\bm \omega_L$ and $\bm
\omega_C$ being the Larmor and cyclotron frequencies
respectively,~see Ref. \cite{glazov04}.

\begin{figure*}[htbp]
  \includegraphics[width=0.9\linewidth]{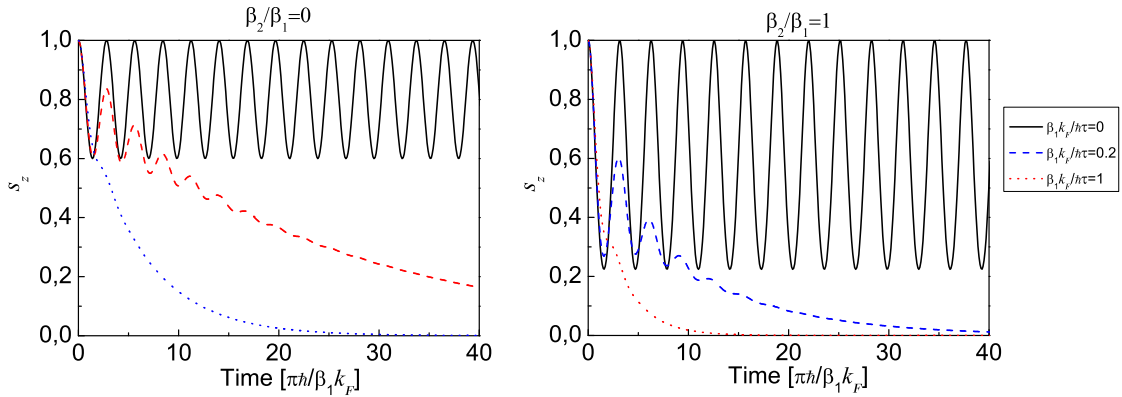}\\
  \caption{The time dynamics of the average $z$-component of the electron
  spin in the magnetic field for different values of the scattering rate
  $\tau^{-1}$, see legend.  Left panel: isotropic spin
  splitting $\beta_2/\beta_1 = 0$, right panel: strongest anisotropy
  $\beta_2/\beta_1=1$.}\label{fig:oscmagn1}
\end{figure*}

First, we start from the clean limit where the scattering is
absent and the spin-splitting is \emph{isotropic}. One can use the
rotating frame of reference where the $\bm \Omega_{\bm k}$ does
not rotate. In this frame the electron spin experiences the
magnetic field being a sum of three components: $\bm \Omega_{\bm
k}$, $\bm \omega_L$ and effective field arising due to transition
to the non-inertial frame which equals to $\pm \bm \omega_C$
(upper sign for $\beta_2\ne 0$ and lower sign for
$\beta_1\ne0$)~\cite{wilamowski}. The spin of the Fermi-surface
electron exhibits the harmonic oscillations around the non-zero
value $\omega^2/(\omega^2+\Omega_{k_F}^2)$ with the amplitude
$\Omega_{k_F}^2/(\omega^2+\Omega_{k_F}^2)$ and frequency
$\sqrt{\omega^2+ \Omega_{k_F}^2}$ where $\omega = \omega_C\pm
\omega_L$ (sign depends on the splitting mechanism in
question)~\cite{glazov06}.

The difference of beats frequencies for BIA and SIA spin
splittings in the magnetic field allows, in principle, to
determine experimentally the symmetry of the spin splitting. We
note that, however, the cyclotron frequency is much larger than
the Larmor one, $\omega_C \gg \omega_L$ in the most conventional
systems and we disregard the Larmor effect from now on.

For the anisotropic spin splitting spin beats become anharmonic.
Applying Eq. \eqref{stott} to the case of \emph{strongest
anisotropy} $\beta_1 = \beta_2 \equiv \beta$ and taking into
account that $\bm k$ rotates with cyclotron frequency we obtain
for the total spin
\begin{equation}\label{stot_aniso_magn}
s_z(t) = s_z(0) \mathrm J_0\left(\frac{8\beta k}{\hbar \omega_C}
\sin{\frac{\omega_C t}{2}}\right).
\end{equation}
One can see that even a small magnetic field restores the strictly
periodic spin dynamics with the period $T = 4\pi /\omega_C$, i.e.
twice larger than the cyclotron one. The time dependence of the
total spin is quite non-trivial: the spin beats demonstrate
multiple harmonics.

In the presence of the scattering, oscillations decay with time as
it is seen from the left panel of Fig.~\ref{fig:oscmagn1}. For the
anisotropic spin splitting the situation is qualitatively the
same, see the right panel of Fig.~\ref{fig:oscmagn1}.

If the magnetic field becomes so strong that $\omega_C \gg \bar
\Omega_{k_F}$ the variations of the effective field $\bm
\Omega_{\bm k}$ occur on the shorter time-scale than the spin
precession in this field. It means that the anisotropic part of
the spin distribution $|\bm s_{\bm k} - \bar {\bm s}_{ k}| \ll
s_{z,k}$. In other words, spin rotation angles between the strong
variations of $\bm\Omega_{\bm k}$ are small. This regime is
analogous to the collision dominated regime where the scattering
is faster than the spin rotation. Thus, Eq. \eqref{kin} can be
solved by iterations and the total spin decays exponentially with
the time-constant
\begin{equation}\label{tau_s_b}
\tau_s = (1+\omega_C^2\tau^2)/\tilde{\Omega}_0^2\tau,
\end{equation}
 where
$\tilde{\Omega}_0$ is given by Eq. \eqref{tildeomega1}. We
underline that Eq. \eqref{tau_s_b} (as well as the results of
Ref.~\cite{glazov04}) is valid for the arbitrary relation between
$\bar \Omega_{k_F}$ and $\tau^{-1}$ provided the magnetic field is
strong enough.

The results presented above are derived for the case of the
isotropic scattering, where all the angular harmonics of the
distribution functions relax with the same time-constant $\tau_m =
\tau$ (for $m=1,2,\ldots$). Qualitatively, these results hold true
for the case of angular dependent scattering. Numerical solution
of Eq.~\eqref{kin} in the limiting case of the long-range
classical potential, where the collision integral can be replaced
by differential operator $\bm Q_{\bm k}\{\bm s_{\bm k} \} =
-\tau^{-1} \partial^2 \bm s_{\bm k}/\partial \varphi_{\bm k}^2$,
shows almost no difference with the presented results.

In conclusion, we have theoretically analyzed the spin dynamics of
the two dimensional electron gas in asymmetrical QWs for the
arbitrary relation between the scattering rate and the spin
precession rate at low temperature regime. We have demonstrated
that the spin dynamics shows quite complicated beats for the
anisotropic spin splitting. In this case the beats decay due to
the spread of the spin precession frequencies. The suppression of
the beats by the thermal smearing of the distribution function,
carrier scattering and the magnetic field has been studied.

\acknowledgements
Author appreciates the valuable discussions with
Profs. N.S. Averkiev, L.E. Golub, R.T. Harley, and E.L. Ivchenko.
The work was partially supported by RFBR, programs of RAS, and
``Dynasty'' foundation -- ICFPM.


\end{document}